\documentclass[fleqn,showpacs,showkeys]{revtex4}
\usepackage{graphicx}
\usepackage{amssymb}
\usepackage{amsmath}
\usepackage{bm}
\usepackage[flushleft]{caption2}

\begin{document}
\setcounter{page}{1}
\title{Octonionic Quantum Interplays of Dark Matter and Ordinary Matter}
\author{Zihua Weng}
\email{xmuwzh@hotmail.com; xmuwzh@xmu.edu.cn.}
\affiliation{School
of Physics and Mechanical \& Electrical Engineering,
\\Xiamen University, Xiamen 361005, China}

\begin{abstract}
Based on the electromagnetic interaction and gravitational
interaction, the quantum interplays of the ordinary matter and the
dark matter in the octonion space are discussed. The paper presents
the quantization of the particles of ordinary matter and dark
matter, including the quantization of the electromagnetic field and
gravitational field etc. In the electromagnetic and gravitational
interactions, it deduces some predictions of the field source
particle, which are consistent with the Dirac equation and
Schrodinger equation in the quantum mechanics. The researches show
that there exist some quantum characteristics of the intermediate
particle in the electromagnetic and gravitational interactions,
including the Dirac-like equation etc.
\end{abstract}



\maketitle

\section{INTRODUCTION}

At the present time, most scientists believe that the universe
abounds with multiform dark matters \cite{bosma, rubin}. A new
viewpoint on the problem of dark matter can be given by the concept
of octonion space. According to previous research results, the
electromagnetic and gravitational interactions can be described by
the quaternion space. Based on the conception of space verticality
etc., two types of quaternion spaces can be united into an octonion
space. In the octonionic space, the electromagnetic and
gravitational interactions can be equally described. So the
characteristics of the field source particle (electron and proton
etc.) and intermediate particle (photon etc.) in the electromagnetic
field, gravitational field, and dark matter field can be described
by the octonion space uniformly \cite{weng}.

Quantum mechanics does not deal with the problem of the quantization
of the dark matter. For the quantization of the gravitational field
and other fields, S. L. Adler \cite{adler} etc. developed some new
forms of quantum mechanics by means of the quaternion, but they did
not consider the case of the dark matter in their researches. And
then the puzzle of the quantization of the dark matter remains
unclear and has not satisfied results.

The paper extends the quantization for the ordinary matter and dark
matter, and draws some conclusions which are consistent with the
Dirac equation, Schrodinger equation, and Dirac-like equation etc. A
few predictions which are associated with the quantum feature of
dark matter can be deduced, and some new and unknown particles can
be used for the candidate of dark matter.

\section{Electromagnetic-gravitational field}

In the electromagnetic-gravitational theory, the physics
characteristics of electromagnetic and gravitational fields can be
described by the octonion space which is united from a couple of
quaternion spaces. And the equations set of the
electromagnetic-gravitational field can be attained.

\subsection{Octonion space}

The base $\mathbb{E}_g$ of the quaternion space of the gravitational
interaction (G space, for short) is $\mathbb{E}_g$ = ($1$,
$\emph{\textbf{i}}_1$, $\emph{\textbf{i}}_2$, $\emph{\textbf{i}}_3$)
. The base $\mathbb{E}_e$ of the quaternion space of the
electromagnetic interaction (E space, for short) is independent of
the base $\mathbb{E}_g$ . Selecting $\mathbb{E}_e$ = ($1$,
$\emph{\textbf{i}}_1$, $\emph{\textbf{i}}_2$, $\emph{\textbf{i}}_3$)
$\circ$ $\emph{\textbf{I}}_0$ = ($\emph{\textbf{I}}_0$,
$\emph{\textbf{I}}_1$, $\emph{\textbf{I}}_2$,
$\emph{\textbf{I}}_3$). So the base $\mathbb{E}_g$ and
$\mathbb{E}_e$ can constitute the base $\mathbb{E}$ of the octonion
space.
\begin{equation}
\mathbb{E} = \mathbb{E}_g + \mathbb{E}_e = (1, \emph{\textbf{i}}_1,
\emph{\textbf{i}}_2, \emph{\textbf{i}}_3, \emph{\textbf{I}}_0,
\emph{\textbf{I}}_1, \emph{\textbf{I}}_2, \emph{\textbf{I}}_3)
\end{equation}

The radius vector $\mathbb{R} (r_0 , r_1 , r_2 , r_3 , R_0 , R_1 ,
R_2 , R_3 )$ in the octonion space is
\begin{equation}
\mathbb{R} = (r_0 + \emph{\textbf{i}}_1 r_1 + \emph{\textbf{i}}_2
r_2 + \emph{\textbf{i}}_3 r_3)+(\emph{\textbf{I}}_0 R_0 +
\emph{\textbf{I}}_1 R_1 + \emph{\textbf{I}}_2 R_2 +
\emph{\textbf{I}}_3 R_3)
\end{equation}
where, $r_0 = v_0 t$, $R_0 = V_0 T$. $v_0 = V_0 = c$ is the speed of
light beam, $t$ and $T$ denote the time.

The octonion algebra is the alternative algebra, so the octonions
$\mathbb{Q}_1$ and $\mathbb{Q}_2$  satisfy
\begin{eqnarray}
\mathbb{Q}_1 \circ ( \mathbb{Q}_1 \circ \mathbb{Q}_2 ) =
(\mathbb{Q}_1 \circ \mathbb{Q}_1) \circ \mathbb{Q}_2~,~~
\mathbb{Q}_1 \circ ( \mathbb{Q}_2 \circ \mathbb{Q}_2 ) =
(\mathbb{Q}_1 \circ \mathbb{Q}_2) \circ \mathbb{Q}_2~. \nonumber
\end{eqnarray}

The octonion differential operator $\lozenge$ and its conjugate
operator $\lozenge^*$ are defined as
\begin{eqnarray}
\lozenge = \lozenge_g + \lozenge_e~.~~\lozenge_g = \partial_{g0} +
\emph{\textbf{i}}_1 \partial_{g1} + \emph{\textbf{i}}_2
\partial_{g2} + \emph{\textbf{i}}_3 \partial_{g3}~;~~\lozenge_e = \emph{\textbf{I}}_0
\partial_{e0} + \emph{\textbf{I}}_1 \partial_{e1} + \emph{\textbf{I}}_2
\partial_{e2} +\emph{\textbf{I}}_3 \partial_{e3}~.
\end{eqnarray}
where, $\partial_{gi}$ = $\partial$/$\partial r_i$ ; $\partial_{ei}$
= $\partial$/$\partial R_i$ ; $i = 0, 1, 2, 3$ .

The field potential $\mathbb{A} (a_0 , a_1 , a_2 , a_3 , A_0 , A_1 ,
A_2 , A_3 )$ in the electromagnetic-gravitational field is defined
as
\begin{eqnarray}
\mathbb{A} = \lozenge^* \circ \mathbb{X} = (a_0 +
\emph{\textbf{i}}_1 a_1 + \emph{\textbf{i}}_2 a_2 +
\emph{\textbf{i}}_3 a_3) + k_a (\emph{\textbf{I}}_0 A_0 +
\emph{\textbf{I}}_1 A_1 + \emph{\textbf{I}}_2 A_2 +
\emph{\textbf{I}}_3 A_3)
\end{eqnarray}
where, $\mathbb{X}=\mathbb{X}_g + k_{rx} \mathbb{X}_e$ ;
$\mathbb{X}_g$ and $\mathbb{X}_e$ are physical quantities in G space
and in E space respectively; $\mathbb{A}=\mathbb{A}_g + k_a
\mathbb{A}_e$; $\mathbb{A}_g = (a_0 , a_1 , a_2 , a_3)$ and
$\mathbb{A}_e = (A_0 , A_1 , A_2 , A_3)$ are the field potential in
G space and in E space respectively; $k_a$ and $k_{rx}$ are the
coefficients.

\begin{table}[t]
\caption{\label{tab:table1}The octonion multiplication table.}
\begin{ruledtabular}
\begin{tabular}{ccccccccc}
$ $ & $1$ & $\emph{\textbf{i}}_1$  & $\emph{\textbf{i}}_2$ &
$\emph{\textbf{i}}_3$  & $\emph{\textbf{I}}_0$  &
$\emph{\textbf{I}}_1$
& $\emph{\textbf{I}}_2$  & $\emph{\textbf{I}}_3$  \\
\hline $1$ & $1$ & $\emph{\textbf{i}}_1$  & $\emph{\textbf{i}}_2$ &
$\emph{\textbf{i}}_3$  & $\emph{\textbf{I}}_0$  &
$\emph{\textbf{I}}_1$
& $\emph{\textbf{I}}_2$  & $\emph{\textbf{I}}_3$  \\
$\emph{\textbf{i}}_1$ & $\emph{\textbf{i}}_1$ & $-1$ &
$\emph{\textbf{i}}_3$  & $-\emph{\textbf{i}}_2$ &
$\emph{\textbf{I}}_1$
& $-\emph{\textbf{I}}_0$ & $-\emph{\textbf{I}}_3$ & $\emph{\textbf{I}}_2$  \\
$\emph{\textbf{i}}_2$ & $\emph{\textbf{i}}_2$ &
$-\emph{\textbf{i}}_3$ & $-1$ & $\emph{\textbf{i}}_1$  &
$\emph{\textbf{I}}_2$  & $\emph{\textbf{I}}_3$
& $-\emph{\textbf{I}}_0$ & $-\emph{\textbf{I}}_1$ \\
$\emph{\textbf{i}}_3$ & $\emph{\textbf{i}}_3$ &
$\emph{\textbf{i}}_2$ & $-\emph{\textbf{i}}_1$ & $-1$ &
$\emph{\textbf{I}}_3$  & $-\emph{\textbf{I}}_2$
& $\emph{\textbf{I}}_1$  & $-\emph{\textbf{I}}_0$ \\
\hline $\emph{\textbf{I}}_0$ & $\emph{\textbf{I}}_0$ &
$-\emph{\textbf{I}}_1$ & $-\emph{\textbf{I}}_2$ &
$-\emph{\textbf{I}}_3$ & $-1$ & $\emph{\textbf{i}}_1$
& $\emph{\textbf{i}}_2$  & $\emph{\textbf{i}}_3$  \\
$\emph{\textbf{I}}_1$ & $\emph{\textbf{I}}_1$ &
$\emph{\textbf{I}}_0$ & $-\emph{\textbf{I}}_3$ &
$\emph{\textbf{I}}_2$  & $-\emph{\textbf{i}}_1$
& $-1$ & $-\emph{\textbf{i}}_3$ & $\emph{\textbf{i}}_2$  \\
$\emph{\textbf{I}}_2$ & $\emph{\textbf{I}}_2$ &
$\emph{\textbf{I}}_3$ & $\emph{\textbf{I}}_0$  &
$-\emph{\textbf{I}}_1$ & $-\emph{\textbf{i}}_2$
& $\emph{\textbf{i}}_3$  & $-1$ & $-\emph{\textbf{i}}_1$ \\
$\emph{\textbf{I}}_3$ & $\emph{\textbf{I}}_3$ &
$-\emph{\textbf{I}}_2$ & $\emph{\textbf{I}}_1$  &
$\emph{\textbf{I}}_0$  & $-\emph{\textbf{i}}_3$
& $-\emph{\textbf{i}}_2$ & $\emph{\textbf{i}}_1$  & $-1$ \\
\end{tabular}
\end{ruledtabular}
\end{table}

\subsection{Dark matter}

In the electromagnetic-gravitational field, there exist four types
of subfields and their field sources. In Table 2, the
electromagnetic-gravitational (E-G) subfield and
gravitational-gravitational (G-G) subfield are 'electromagnetic
field' and 'gravitational field' respectively. And their general
charges are electronic charge (G charge) and the mass (G mass)
respectively. The electromagnetic-electromagnetic (E-E) and
gravitational-electromagnetic (G-E) subfields are both long range
fields and candidates of the 'dark matter field'. Their general
charges (E charge and E mass) are candidates of 'dark matter'. The
physical features of the dark matter meet the requirements of
Eqs.(4)$-$(10).

\begin{table*}[t]
\begin{ruledtabular}
\caption{\label{tab:table1}The subfield types of
electromagnetic-gravitational field.}
\begin{tabular}{lll}
\textbf{}Operator      & Gravitational Interaction               & Electromagnetic Interaction\\
\hline
operator $\lozenge_g$  & gravitational-gravitational subfield    & electromagnetic-gravitational subfield\\
of G space             & (gravitational field, or G-G subfield)  & (electromagnetic field, or E-G subfield)\\
                       & G mass, $Q_g^g$                         & G charge, $Q_e^g$ \\
                       & intermediate particle, $\gamma_g^g$     & intermediate particle, $\gamma_e^g$\\
                       & long range field, weak strength         & long range field, strong strength\\
\hline
operator $\lozenge_e$  & gravitational-electromagnetic subfield  & electromagnetic-electromagnetic subfield\\
of E space             & (dark matter field, or G-E subfield)    & (dark matter field, or E-E subfield)\\
                       & E mass, $Q_g^e$                         & E charge, $Q_e^e$ \\
                       & intermediate particle, $\gamma_g^e$     & intermediate particle, $\gamma_e^e$\\
                       & long range field, weaker strength       & long range field, weaker strength\\
\end{tabular}
\end{ruledtabular}
\label{tab:front}
\end{table*}

In the general charge $(Q_e^e , Q_e^g , Q_g^e , Q_g^g )$ and
intermediate particle $( \gamma_e^e , \gamma_e^g , \gamma_g^e ,
\gamma_g^g )$, two types of general charges (the electronic charge
$Q_e^g$, the mass $Q_g^g$) and one type of intermediate particle
(the photon $\gamma_e^g$) have been found. Therefore the two types
of general charges and three types of intermediate particles are
left to be found in Table 2.

The particles of the ordinary matter (the electron and proton etc.)
possess the G charge together with G mass. The particles of the dark
matter may possess the E charge with E mass, or G mass with E
charge, etc. It can be predicted that the field strength of the
electromagnetic-electromagnetic and the
gravitational-electromagnetic subfields must be weaker than that of
the gravitational-gravitational subfield, otherwise they should be
detected for a long time. So the field strength of the
electromagnetic-electromagnetic subfield and
gravitational-electromagnetic subfield may be equal, and both of
them are slightly weaker than that of the
gravitational-gravitational subfield.

\begin{table*}[b]
\caption{\label{tab:table3} The comparison between the ordinary
matter fields with dark matter fields.}
\begin{ruledtabular}
\begin{tabular}{ccccc}
&\multicolumn{2}{c}{Ordinary~Matter~Fields}&\multicolumn{2}{c}{Dark~Matter~Fields}\\
 \hline
(field) & (gravitational field) & (electromagnetic field) & (dark
 matter field) & (dark matter field)\\
 subfield & G-G~subfield & E-G~subfield & G-E~subfield & E-E~subfield \\
 field~potential & $\lozenge_g^* \circ \mathbb{X}_g$ & $\lozenge_g^* \circ \mathbb{X}_e$
 & $\lozenge_e^* \circ \mathbb{X}_g$ & $\lozenge_e^* \circ \mathbb{X}_e$\\
 field~strength & $\lozenge_g \circ \mathbb{A}_g$ & $\lozenge_g \circ \mathbb{A}_e$
 & $\lozenge_e \circ \mathbb{A}_g$ & $\lozenge_e \circ \mathbb{A}_e$\\
 field~source & $\lozenge_g^* \circ \mathbb{B}_g$ & $\lozenge_g^* \circ \mathbb{B}_e$
 & $\lozenge_e^* \circ \mathbb{B}_g$ & $\lozenge_e^* \circ \mathbb{B}_e$\\
\end{tabular}
\end{ruledtabular}
\end{table*}

\subsection{Equations sets}

In the electromagnetic-gravitational field, the force $\mathbb{Z}$
and energy $\mathbb{W}$ can be defined uniformly, and the octonion
differential operator $\lozenge$ needs to be generalized to the
$(\lozenge + \mathbb{B}/\alpha)$. So the physical characteristics of
the electromagnetic-gravitational field can be researched from many
aspects.

The field strength $\mathbb{B} ( b_0 , b_1 , b_2 , b_3 , B_0 , B_1 ,
B_2 , B_3 )$ of electromagnetic-gravitational field can be defined
as
\begin{eqnarray}
\mathbb{B}=  \lozenge \circ \mathbb{A}
\end{eqnarray}
where, $\mathbb{B} = \mathbb{B}_g + k_b \mathbb{B}_e$ ;
$\mathbb{B}_g = (b_1 , b_2 , b_3)$ and $\mathbb{B}_e = (B_1 , B_2 ,
B_3)$ are respectively the field strength in G space and in E space.
Selecting the gauge equation, $b_0 = 0$ and $B_0 = 0$, can simplify
definition of field strength.

The field source and the force of the electromagnetic-gravitational
field can be defined as follow respectively. The mark (*) denotes
octonion conjugate. ($\alpha = c$ is a coefficient )
\begin{eqnarray}
&& \mu \mathbb{S}= (\mathbb{B}/\alpha + \lozenge)^* \circ \mathbb{B}
\\
&& \mathbb{Z}= \alpha (\mathbb{B}/\alpha + \lozenge)^* \circ
\mathbb{P}
\end{eqnarray}

The definition of force shows that, the force $\mathbb{Z}$ , field
source $\mathbb{S}$ , and linear momentum $\mathbb{P} = \mu
\mathbb{S}/\mu_g^g$ need to be revised and generalized to the form
in above equations. As a part of field source $\mathbb{S}$, the term
$(\mathbb{B}^* \circ \mathbb{B}/\mu_g^g)$ includes the field energy
density. When the force $\mathbb{Z} = 0$, we obtain the
force-balance equation of the electromagnetic-gravitational field.

The angular momentum in the electromagnetic-gravitational field can
be defined as,
\begin{eqnarray}
\mathbb{M}= (\mathbb{R} + k_{rx} \mathbb{X}) \circ \mathbb{P}
\end{eqnarray}
where, $k_{rx}$ is a coefficient .

And the energy and power in the electromagnetic-gravitational field
can be defined as
\begin{eqnarray}
&& \mathbb{W}= \alpha (\mathbb{B}/\alpha + \lozenge)^* \circ
\mathbb{M}
\\
&& \mathbb{N}= \alpha (\mathbb{B}/\alpha + \lozenge)^* \circ
\mathbb{W}
\end{eqnarray}

The extended definition of the energy shows that, the angular
momentum $\mathbb{M}$, the energy $\mathbb{W}$ and the power
$\mathbb{N}$ need to be revised and generalized to the form in the
above equations. The physical quantity $\mathbb{X}$ has effect on
the field potential $\mathbb{A} = \lozenge^* \circ \mathbb{X}$ , the
angular momentum $\mathbb{X} \circ \mathbb{P}$ , the energy
$\mathbb{B}^* \circ (\mathbb{X} \circ \mathbb{P})$ and the power
$(\mathbb{B} \circ \mathbb{B}^*) \circ (\mathbb{X} \circ
\mathbb{P})$ . The introduction of physical quantity $\mathbb{X}$
makes the definition of the angular momentum $\mathbb{M}$ and the
energy $\mathbb{W}$ more integrated, and the theory more
self-consistent.

In the above equations, the conservation of angular momentum in the
electromagnetic-gravitational field can be gained when $\mathbb{W} =
0$, and the energy conservation equation in the
electromagnetic-gravitational field can be attained when $\mathbb{N}
= 0$.

\section{Equations of quantum mechanics}

\subsection{Dirac Equation}

In the octonion space, the wave functions of the quantum mechanics
are the octonion equations set. And the Dirac equation of the
quantum mechanics are actually the wave equations set which are
associated with the particle's wave function $- \emph{\textbf{I}}
\circ \mathbb{M}/\hbar$ .

The $\mathbb{U}$ equation of the quantum mechanics can be defined as
\begin{eqnarray}
\mathbb{U}= (\mathbb{W}/\alpha + \hbar \lozenge)^* \circ (-
\emph{\textbf{I}} \circ \mathbb{M}/\hbar)
\end{eqnarray}
where, $- \emph{\textbf{I}} \circ \mathbb{M}/\hbar$ is the wave
function for particles of the field source; $\emph{\textbf{I}}$ is
the octonion unit, $\emph{\textbf{I}}^* \circ \emph{\textbf{I}} =
1$; the coefficient $h$ is the Planck constant, and $ \hbar = h/2\pi
$ .

The $\mathbb{L}$ equation of the quantum mechanics can be defined as
\begin{eqnarray}
\mathbb{L}= (\mathbb{W}/\alpha + \hbar \lozenge)^* \circ
(\mathbb{U}/\hbar)
\end{eqnarray}

From the equation $\mathbb{U} = 0$, the Dirac and Schrodinger
equations in the electromagnetic-gravitational field can be deduced
to describe the field source particle (electron and proton etc.)
with spin $1/2$ . Those equations can conclude the results which are
consistent with the wave equations for particles of the field source
in certain cases \cite{wu, boulanger, singh}.

\subsection{Dirac-like Equation}

Through the comparison, we find that Dirac equation of Eq.(11) and
Eq.(12) can be attained respectively from the energy equation Eq.(9)
and power equation Eq.(10) after substituting the operator $\alpha
(\mathbb{B}/\alpha + \lozenge)$ for $\left\{ \mathbb{W}/(\alpha
\hbar ) + \lozenge \right\}$ . By analogy with the above equations,
the Dirac-like equations of Eq.(13) and (14) can be obtained from
the field source equation Eq.(6) and force equation Eq.(7)
respectively.

The $\mathbb{T}$ equation of the quantum mechanics can be defined as
\begin{eqnarray}
\mathbb{T}= (\mathbb{W}/\alpha + \hbar \lozenge)^* \circ
(\mathbb{B}/\hbar)
\end{eqnarray}
where, $\mathbb{B}/\hbar$ is the wave function for the intermediate
particle.

The $\mathbb{O}$ equation of the quantum mechanics can be defined as
\begin{eqnarray}
\mathbb{O}= (\mathbb{W}/\alpha + \hbar \lozenge)^* \circ
(\mathbb{T}/\hbar)
\end{eqnarray}

From the equation $\mathbb{T} = 0$, the Dirac-like equation in
electromagnetic-gravitational field can be deduced to describe the
intermediate particle (photon etc.) with spin 1 . Those equations
can conclude results which are consistent with the wave function for
the intermediate particle in certain cases \cite{lomont, moses}.

\begin{table*}[h]
\begin{ruledtabular}
\caption{\label{tab:table1}The summary of main definitions and
equations.}
\begin{tabular}{ll}
\textbf{} $\mathbb{X}$ physical quantity      & $\mathbb{X}$ \\
Field potential          &  $\mathbb{A} = \lozenge^* \circ \mathbb{X}$ \\
Field strength           &  $\mathbb{B} = \lozenge \circ \mathbb{A}$ \\
Field source             &  $\mu \mathbb{S} = (\mathbb{B}/\alpha + \lozenge)^* \circ \mathbb{B}$ \\
Force                    &  $\mathbb{Z} = \alpha (\mathbb{B}/\alpha + \lozenge)^* \circ \mathbb{P}$ \\
\hline
Angular momentum         &  $\mathbb{M} = (\mathbb{R} + k_{rx} \mathbb{X}) \circ \mathbb{P}$ \\
Energy                   &  $\mathbb{W} = \alpha (\mathbb{B}/\alpha + \lozenge)^* \circ \mathbb{M}$ \\
Power                    &  $\mathbb{N} = \alpha (\mathbb{B}/\alpha + \lozenge)^* \circ \mathbb{W}$ \\
\hline Energy quantum    &  $\mathbb{U} = (\mathbb{W}/\alpha
                             + \hbar \lozenge)^* \circ (- \emph{\textbf{I}} \circ \mathbb{M}/\hbar)$ \\
Power quantum            &  $\mathbb{L} = (\mathbb{W}/\alpha + \hbar \lozenge)^* \circ (\mathbb{U}/\hbar)$ \\
Field source quantum     &  $\mathbb{T} = (\mathbb{W}/\alpha + \hbar \lozenge)^* \circ (\mathbb{B}/\hbar)$ \\
Force quantum            &  $\mathbb{O} = (\mathbb{W}/\alpha + \hbar \lozenge)^* \circ (\mathbb{T}/\hbar)$ \\
\end{tabular}
\end{ruledtabular}
\label{tab:front}
\end{table*}

\subsection{Wave function}

The wave function $- \emph{\textbf{I}} \circ \mathbb{M}/\hbar$ is
that of the field source particle (electron and proton etc.), and
its square module represents the probability density of field source
particle. The wave function $\mathbb{B}/\hbar$ is that of the wave
function of intermediate particle (photon etc.), and its square
module represents the probability density of intermediate particle.

The octonion physical quantity $\mathbb{Q} (q_0 , q_1 , q_2 , q_3 ,
Q_0 , Q_1 , Q_2 , Q_3)$ can be written as exponential format
\begin{eqnarray}
\mathbb{Q} && = (q_0 + \emph{\textbf{i}}_1 q_1 + \emph{\textbf{i}}_2
q_2 + \emph{\textbf{i}}_3 q_3)+(\emph{\textbf{I}}_0 Q_0 +
\emph{\textbf{I}}_1 Q_1 + \emph{\textbf{I}}_2 Q_2 +
\emph{\textbf{I}}_3 Q_3) \nonumber
\\
&& = (q_0 + \emph{\textbf{i}}_1 q_1 + \emph{\textbf{i}}_2 q_2 +
\emph{\textbf{i}}_3 q_3)+(\emph{\textbf{i}}_0 Q_0 +
\emph{\textbf{i}}_1 Q_1 + \emph{\textbf{i}}_2 Q_2 +
\emph{\textbf{i}}_3 Q_3) \circ \emph{\textbf{I}}_0 \nonumber
\\
&& = q ~exp(\emph{\textbf{q}}_1 \omega_1) + Q
~exp(\emph{\textbf{q}}_2 \omega_2) \circ \emph{\textbf{I}}_0
\end{eqnarray}
where, $q$ and $Q$ are the modules; $\omega_1$ and $\omega_2$ are
the angles; $\emph{\textbf{q}}_1$ and $\emph{\textbf{q}}_2$ are the
unit vectors in the quaternion G space.

In the quaternion space, the quaternion product of the coefficient
$1/\hbar $ , momentum $\mathbb{P} (p_0 , p_1 , p_2 , p_3 )$ and
radius vector $\mathbb{R} ( r_0 , r_1 , r_2 , r_3 )$ is
\begin{eqnarray}
\mathbb{M}/\hbar = \mathbb{R} \circ \mathbb{P}/\hbar = A (cos\theta
+ \emph{\textbf{I}} sin\theta) = A~exp(\emph{\textbf{I}}\theta)
\end{eqnarray}
where, $\emph{\textbf{I}}$ is the quaternion unit,
$\emph{\textbf{I}}^* \circ \emph{\textbf{I}} = 1$ , $A$ is the
amplitude; $\mathbb{X} \approx 0$ .

The scalar quantity $\nu_0$ of the quaternion $\mathbb{M}/\hbar$ is
\begin{eqnarray}
\nu_0 = A~cos\theta = (p_0 r_0 - p_1 r_1 - p_2 r_2 - p_3 r_3)/\hbar
\end{eqnarray}
where, $p_0 = mc$ , $r_0 = ct$.

When $\nu_0 /A \ll 1$, the angle can be written as the expansion of
Taylor progression
\begin{eqnarray}
\theta = arccos (\nu_0 /A) \approx \pi/2 - \nu_0 /A
\end{eqnarray}
therefore we have the wave function
\begin{eqnarray}
\mathbb{M}/\hbar = \mathbb{R} \circ \mathbb{P}/\hbar =
A~\emph{\textbf{I}} \circ exp(-\emph{\textbf{I}}~\nu_0 /A)
\end{eqnarray}

The quaternion wave function can also be defined as
\begin{eqnarray}
\Psi = - \emph{\textbf{I}} \circ (\mathbb{M}/\hbar) = A~
exp\left\{-\emph{\textbf{I}}~(p_0 r_0 - p_1 r_1 - p_2 r_2 - p_3
r_3)/\hbar\right\}
\end{eqnarray}

The above means that the matter can be represented as either the
particle or the wave in the quaternion space or octonion space. In
certain cases, the quaternion wave function can be written as the
four-component or exponential format. If its direction could be
neglected, the quaternion unit $\emph{\textbf{I}}$ would be
substituted with the imaginary unit $i$. And a couple of quaternion
wave functions can be written as the eight-component or exponential
format in the octonion space in the same way as Eq.(15).

\section{Quantization of ordinary matter}

In the electromagnetic-gravitational field, there is only one sort
of the ordinary matter with a pair of general charges. Their Dirac
and Schrodinger equations can be attained when $\mathbb{U} = 0$ from
Eq.(11). And the Dirac-like equation can be obtained when
$\mathbb{T} = \mathbb{W} = 0$ from Eq.(13). With the characteristics
of those equations, we can investigate the quantum properties of the
field source particle and intermediate particle.

\subsection{Dirac and Schrodinger equations}

In the octonion space, the electromagnetic-gravitational subfield
(electromagnetic field) and gravitational-gravitational subfield
(gravitational field) are generated by the physical object $M$ which
owns rotation and charge. The G current and G momentum of the field
source particle $N(m,q)$ are $(S_0^g , S_1^g , S_2^g , S_3^g)$ and
$(s_0^g , s_1^g , s_2^g , s_3^g)$ respectively. The $N(m,q)$ is the
mixture of the general charges $Q_g^g$ (the mass, $m$) and $Q_e^g$
(the electric charge, $q$). When $\mathbb{U} = 0$ , the wave
equation of the particle $N(m,q)$ which moves around $M$ is
\begin{eqnarray}
(\mathbb{W}/\alpha + \hbar \lozenge)^* \circ (- \emph{\textbf{I}}
\circ \mathbb{M}/\hbar) = 0
\end{eqnarray}

Because of $ | S_0^g | \gg |  S_i^g | $ and $ | s_0^g | \gg | s_i^g
| $ , then $(i = 1, 2, 3)$
\begin{eqnarray}
\mathbb{W} && = (\mathbb{B} + \alpha \lozenge)^* \circ \left\{ (
\mathbb{R} + k_{rx} \mathbb{X} ) \circ \mathbb{P} \right\}
\nonumber\\
&& \approx \alpha \lozenge^* \circ \left\{ ( \mathbb{R} + k_{rx}
\mathbb{X} ) \circ (\mu_g^g s_0^g + k_b \mu_e^g \emph{\textbf{I}}_0
S_0^g) \right\}/\mu_g^g + \mathbb{B}^* \circ \mathbb{M}
\nonumber\\
&& \approx \alpha (m \mathbb{V} + q \mathbb{A}') + \mathbb{B}^*
\circ \mathbb{M} + \alpha k_b \mu_e^g S_0^g \left\{ \lozenge^* \circ
( \mathbb{R} \circ \emph{\textbf{I}}_0 )\right\}/ \mu_g^g  + k_{rx}
s_0^g \mathbb{A} \nonumber
\end{eqnarray}
where, $\mathbb{A}' = (c k_{rx} k_b \mu_e^g/\mu_g^g) \lozenge^*
\circ (\mathbb{X} \circ \emph{\textbf{I}}_0 ) = a'_0 +
\emph{\textbf{i}}_1 a'_1 + \emph{\textbf{i}}_2 a'_2 +
\emph{\textbf{i}}_3 a'_3 + \emph{\textbf{I}}_0 A'_0 +
\emph{\textbf{I}}_1 A'_1 + \emph{\textbf{I}}_2 A'_2 +
\emph{\textbf{I}}_3 A'_3 $; $S_0^g = qc , s_0^g = mc$; $\mathbb{V} =
v_0 + \emph{\textbf{i}}_1 v_1 + \emph{\textbf{i}}_2 v_2 +
\emph{\textbf{i}}_3 v_3 + \emph{\textbf{I}}_0 V_0 +
\emph{\textbf{I}}_1 V_1 + \emph{\textbf{I}}_2 V_2 +
\emph{\textbf{I}}_3 V_3 $ ; $m$ is the $Q_g^g$, $q$ is the $Q_e^g$ .

When the $\mathbb{B}$ is small, and the sum of last three terms is
equal approximately to zero, the above equation can be written as
follows
\begin{eqnarray}
\mathbb{W}/\alpha = && m \mathbb{V} + q \mathbb{A}'
\nonumber\\
= && (qa'_0 + mv_0) + \emph{\textbf{i}}_1 (qa'_1 + mv_1) +
\emph{\textbf{i}}_2 (qa'_2 + mv_2) + \emph{\textbf{i}}_3 (qa'_3 +
mv_3)
\nonumber\\
&& + \emph{\textbf{I}}_0 (qA'_0 + mV_0) + \emph{\textbf{I}}_1 (qA'_1
+ mV_1) + \emph{\textbf{I}}_2 (qA'_2 + mV_2) + \emph{\textbf{I}}_3
(qA'_3 + mV_3)
\nonumber\\
= && p_0 + \emph{\textbf{i}}_1 p_1 + \emph{\textbf{i}}_2 p_2 +
\emph{\textbf{i}}_3 p_3 + \emph{\textbf{I}}_0 P_0 +
\emph{\textbf{I}}_1 P_1 + \emph{\textbf{I}}_2 P_2 +
\emph{\textbf{I}}_3 P_3
\end{eqnarray}
where, $p_j = qa'_j + mv_j ~; P_j = qA'_j + mV_j ~; j = 0, 1, 2, 3$
.

Then
\begin{eqnarray}
0 = && ( \mathbb{W}/\alpha + \hbar \lozenge)^* \circ \left\{ (
\mathbb{W}/\alpha + \hbar \lozenge)^*  \circ \Psi \right\}
\nonumber\\
\approx &&  [ (p_0 + \hbar \partial_{g0})^2 - (p_1 + \hbar
\partial_{g1})^2 - (p_2 + \hbar \partial_{g2})^2 - (p_3 + \hbar \partial_{g3})^2
- (P_0 + \hbar \partial_{e0})^2
\nonumber\\
&& - (P_1 + \hbar \partial_{e1})^2 - (P_2 + \hbar \partial_{e2})^2 -
(P_3 + \hbar \partial_{e3})^2 + q \hbar \lozenge^* \circ
\mathbb{A}'^* + m \hbar \lozenge^* \circ ( \emph{\textbf{I}}_0 \circ
\mathbb{V})^* ] \circ \Psi
\end{eqnarray}
where, $ \Psi = - \emph{\textbf{I}} \circ \mathbb{M} /\hbar $ is the
wave function.

In the above equation, the conservation of wave function is
influenced by the field potential, field strength, field source,
velocity, charge and mass etc. When the linear momentum $P_j$ is
equal approximately to 0, and the wave function is $ \Psi = \psi(r)
exp(-\emph{\textbf{I}}~ E t/\hbar )$, the above equation can be
simplified as
\begin{eqnarray}
0 = &&  [ (p_0 - \emph{\textbf{I}}~ E/c)^2 - (p_1)^2 - (p_2)^2 -
(p_3)^2 + q \hbar \lozenge^* \circ \mathbb{A}'^* ] \circ \psi (r)
\nonumber\\
\approx && [ (qa'_0 - \emph{\textbf{I}}~ E/c) - (1/2mc) \left\{
(p_1)^2 + (p_2)^2 + (p_3)^2 \right\} + (q\hbar/2mc) (\lozenge^*
\circ \mathbb{A}'^*) ] \circ \psi (r)
\end{eqnarray}
where, $E$ is the energy; $(q \hbar/2m)(\lozenge^* \circ
\mathbb{A}'^*)$ is the interplay term of the
electromagnetic-gravitational subfield with the spin $(q \hbar/2m)$.

Limited within certain conditions, Eq.(11) of the
electromagnetic-gravitational field in the octonion space can draw
some conclusions which are consistent with Dirac and Schrodinger
equations about ordinary matter, including the spin and magnetic
moment etc.

\subsection{Intermediate particle equations}

In the octonion space, the electromagnetic-gravitational subfield
and gravitational-gravitational subfield are generated by the
physical object $M$ which owns rotation and charge. The G current
and G momentum of the intermediate particle $N(m,q)$ are $(S_0^g ,
S_1^g , S_2^g , S_3^g)$ and $(s_0^g , s_1^g , s_2^g , s_3^g)$
respectively. The $N(m,q)$ is the mixture of the intermediate
particles $\gamma_e^g$ (the photon) and $\gamma_g^g$ . When
$\mathbb{T} = 0$ , the wave equation of the particle $N(m,q)$ which
moves around $M$ is
\begin{eqnarray}
(\mathbb{W}/\alpha + \hbar \lozenge)^* \circ (\mathbb{B}/\hbar) = 0
\end{eqnarray}

When the $\mathbb{B}$ is small, and the sum of last three terms is
equal approximately to zero, the above equation can be written as
follows
\begin{eqnarray}
\mathbb{W}/\alpha = && m \mathbb{V} + q \mathbb{A}'
\nonumber\\
= && (qa'_0 + mv_0) + \emph{\textbf{i}}_1 (qa'_1 + mv_1) +
\emph{\textbf{i}}_2 (qa'_2 + mv_2) + \emph{\textbf{i}}_3 (qa'_3 +
mv_3)
\nonumber\\
&& + \emph{\textbf{I}}_0 (qA'_0 + mV_0) + \emph{\textbf{I}}_1 (qA'_1
+ mV_1) + \emph{\textbf{I}}_2 (qA'_2 + mV_2) + \emph{\textbf{I}}_3
(qA'_3 + mV_3)
\nonumber\\
= && p_0 + \emph{\textbf{i}}_1 p_1 + \emph{\textbf{i}}_2 p_2 +
\emph{\textbf{i}}_3 p_3 + \emph{\textbf{I}}_0 P_0 +
\emph{\textbf{I}}_1 P_1 + \emph{\textbf{I}}_2 P_2 +
\emph{\textbf{I}}_3 P_3 \nonumber
\end{eqnarray}
where, $\mathbb{A}' = (c k_{rx} k_b \mu_e^g/\mu_g^g) \lozenge^*
\circ (\mathbb{X} \circ \emph{\textbf{I}}_0 ) = a'_0 +
\emph{\textbf{i}}_1 a'_1 + \emph{\textbf{i}}_2 a'_2 +
\emph{\textbf{i}}_3 a'_3 + \emph{\textbf{I}}_0 A'_0 +
\emph{\textbf{I}}_1 A'_1 + \emph{\textbf{I}}_2 A'_2 +
\emph{\textbf{I}}_3 A'_3 $; $S_0^g = qc , s_0^g = mc$; $\mathbb{V} =
v_0 + \emph{\textbf{i}}_1 v_1 + \emph{\textbf{i}}_2 v_2 +
\emph{\textbf{i}}_3 v_3 + \emph{\textbf{I}}_0 V_0 +
\emph{\textbf{I}}_1 V_1 + \emph{\textbf{I}}_2 V_2 +
\emph{\textbf{I}}_3 V_3 $ ; $p_j = qa'_j + mv_j ~; P_j = qA'_j +
mV_j ~; j = 0, 1, 2, 3$.

Then
\begin{eqnarray}
0 = && ( \mathbb{W}/\alpha + \hbar \lozenge)^* \circ \left\{ (
\mathbb{W}/\alpha + \hbar \lozenge)^*  \circ \Psi \right\}
\nonumber\\
\approx &&  [ (p_0 + \hbar \partial_{g0})^2 - (p_1 + \hbar
\partial_{g1})^2 - (p_2 + \hbar \partial_{g2})^2 - (p_3 + \hbar \partial_{g3})^2
- (P_0 + \hbar \partial_{e0})^2
\nonumber\\
&& - (P_1 + \hbar \partial_{e1})^2 - (P_2 + \hbar \partial_{e2})^2 -
(P_3 + \hbar \partial_{e3})^2 + q \hbar \lozenge^* \circ
\mathbb{A}'^* + m \hbar \lozenge^* \circ ( \emph{\textbf{I}}_0 \circ
\mathbb{V})^* ] \circ \Psi
\end{eqnarray}
where, $\Psi = \mathbb{B} /\hbar$ is the wave function; $(q\hbar
/m)(\lozenge^* \circ \mathbb{A}'^*)$ is the interplay term of the
electromagnetic-gravitational subfield with the spin $(q\hbar/m)$.

The above equation can be used to describe the quantum
characteristics of intermediate particles which possess the spin 1,
G charge and G mass. Limited within certain conditions, Eq.(13) of
the electromagnetic-gravitational field in the octonion space can
deduce the wave equation of ordinary matter and its conclusions.

\subsection{Dirac-like equation}

In the octonion space, the electromagnetic-gravitational subfield
and gravitational-gravitational subfield are generated by the
physical object $M$ which owns rotation and charge. The G current
and G momentum of the intermediate particle $N(m,q)$ are $(S_0^g ,
S_1^g , S_2^g , S_3^g)$ and $(s_0^g , s_1^g , s_2^g , s_3^g)$
respectively. The $N(m,q)$ is the mixture of the intermediate
particles $\gamma_g^g$ and $\gamma_e^g$ . When $\mathbb{T} = 0$ ,
the wave equation of the particle $N(m,q)$ which moves around $M$ is
\begin{eqnarray}
(\mathbb{W}/\alpha + \hbar \lozenge)^* \circ (\mathbb{B}/\hbar) = 0
\end{eqnarray}

When the energy $\mathbb{W} = 0$, the Dirac-like equation can be
attained from the above equation
\begin{eqnarray}
\hbar \lozenge^* \circ (\mathbb{B}/\hbar) = 0
\end{eqnarray}

Dirac equation can conclude that field source particles (electron
and proton etc.) possess the spin $(q\hbar/2m)$. In the same way,
the above equation can infer that intermediate particles (photon
etc.) own the spin $(q\hbar/m)$ and have no G charge nor G mass. And
the familiar quantum theory of electromagnetic field can be obtained
\cite{corvino, oaknin, galley}.

Limited within certain conditions, Eq.(13) of the
electromagnetic-gravitational field in the octonion space can draw
some conclusions which are consistent with Dirac-like equation about
ordinary matter.

\section{Quantization of dark matter A}

In the electromagnetic-gravitational field, there are five sorts of
dark matters with a pair of general charges. Two of them will be
discussed here. With the characteristics of Eqs.(11) and (13), we
can investigate the quantum interplays among the
electromagnetic-gravitational subfield (ordinary matter field) with
electromagnetic-electromagnetic subfield (dark matter field), and
describe their quantum properties of the field source particle and
intermediate particle.

\subsection{Dirac and Schrodinger equations}

In the octonion space, the electromagnetic-gravitational subfield
and electromagnetic-electromagnetic subfield are generated by the
physical object $M$ which owns rotation and charge. The E current
and G current of the field source particle $N(d,q)$ are $(S_0^e ,
S_1^e , S_2^e , S_3^e)$ and $(S_0^g , S_1^g , S_2^g , S_3^g)$
respectively. The $N(d,q)$ is the mixture of the general charges
$Q_e^e$ (the dark matter mass, $d$) and $Q_e^g$ (the electric
charge, $q$). When $\mathbb{U} = 0$, the wave equation of particle
$N(d,q)$ which moves around $M$ is
\begin{eqnarray}
(\mathbb{W}/\alpha + \hbar \lozenge)^* \circ (- \emph{\textbf{I}}
\circ \mathbb{M}/\hbar) = 0
\end{eqnarray}

Because of $ | S_0^g | \gg |  S_i^g | $ and $ | S_0^e | \gg | S_i^e
| $ , then $(i = 1, 2, 3)$
\begin{eqnarray}
\mathbb{W} && = (\mathbb{B} + \alpha \lozenge)^* \circ \left\{ (
\mathbb{R} + k_{rx} \mathbb{X} ) \circ \mathbb{P} \right\}
\nonumber\\
&& \approx \alpha \lozenge^* \circ \left\{ ( \mathbb{R} + k_{rx}
\mathbb{X} ) \circ (k_b \mu_e^e S_0^e + k_b \mu_e^g
\emph{\textbf{I}}_0 S_0^g) \right\}/\mu_g^g + \mathbb{B}^* \circ
\mathbb{M}
\nonumber\\
&& \approx \alpha (d \mathbb{V}' + q \mathbb{A}') + \mathbb{B}^*
\circ \mathbb{M} + \alpha k_b [ \mu_e^g S_0^g \left\{ \lozenge^*
\circ ( \mathbb{R} \circ \emph{\textbf{I}}_0 )\right\} + \mu_e^e
k_{rx} S_0^e \mathbb{A}] / \mu_g^g \nonumber
\end{eqnarray}
where, $\mathbb{A}' = (c k_{rx} k_b \mu_e^g/\mu_g^g) \lozenge^*
\circ (\mathbb{X} \circ \emph{\textbf{I}}_0 ) = a'_0 +
\emph{\textbf{i}}_1 a'_1 + \emph{\textbf{i}}_2 a'_2 +
\emph{\textbf{i}}_3 a'_3 + \emph{\textbf{I}}_0 A'_0 +
\emph{\textbf{I}}_1 A'_1 + \emph{\textbf{I}}_2 A'_2 +
\emph{\textbf{I}}_3 A'_3 $; $S_0^g = qc , S_0^e = dc$; $\mathbb{V}'
= (k_b \mu_e^e/\mu_g^g) \mathbb{V} = v'_0 + \emph{\textbf{i}}_1 v'_1
+ \emph{\textbf{i}}_2 v'_2 + \emph{\textbf{i}}_3 v'_3 +
\emph{\textbf{I}}_0 V'_0 + \emph{\textbf{I}}_1 V'_1 +
\emph{\textbf{I}}_2 V'_2 + \emph{\textbf{I}}_3 V'_3 $ ; $q$ is the
$Q_e^g$, $d$ is the $Q_e^e$ .

When the $\mathbb{B}$ is small, and the sum of last two terms is
equal approximately to zero, the above equation can be written as
follows
\begin{eqnarray}
\mathbb{W}/\alpha = && d \mathbb{V}' + q \mathbb{A}'
\nonumber\\
= && (qa'_0 + dv'_0) + \emph{\textbf{i}}_1 (qa'_1 + dv'_1) +
\emph{\textbf{i}}_2 (qa'_2 + dv'_2) + \emph{\textbf{i}}_3 (qa'_3 +
dv'_3)
\nonumber\\
&& + \emph{\textbf{I}}_0 (qA'_0 + dV'_0) + \emph{\textbf{I}}_1
(qA'_1 + dV'_1) + \emph{\textbf{I}}_2 (qA'_2 + dV'_2) +
\emph{\textbf{I}}_3 (qA'_3 + dV'_3)
\nonumber\\
= && p_0 + \emph{\textbf{i}}_1 p_1 + \emph{\textbf{i}}_2 p_2 +
\emph{\textbf{i}}_3 p_3 + \emph{\textbf{I}}_0 P_0 +
\emph{\textbf{I}}_1 P_1 + \emph{\textbf{I}}_2 P_2 +
\emph{\textbf{I}}_3 P_3 \nonumber
\end{eqnarray}
where, $p_j = qa'_j + dv'_j ~; P_j = qA'_j + dV'_j ~; j = 0, 1, 2,
3$.

Therefore
\begin{eqnarray}
0 = && ( \mathbb{W}/\alpha + \hbar \lozenge)^* \circ \left\{ (
\mathbb{W}/\alpha + \hbar \lozenge)^*  \circ \Psi \right\}
\nonumber\\
\approx &&  [ (p_0 + \hbar \partial_{g0})^2 - (p_1 + \hbar
\partial_{g1})^2 - (p_2 + \hbar \partial_{g2})^2 - (p_3 + \hbar \partial_{g3})^2
- (P_0 + \hbar \partial_{e0})^2
\nonumber\\
&& - (P_1 + \hbar \partial_{e1})^2 - (P_2 + \hbar \partial_{e2})^2 -
(P_3 + \hbar \partial_{e3})^2 + q \hbar \lozenge^* \circ
\mathbb{A}'^* + d \hbar \lozenge^* \circ ( \emph{\textbf{I}}_0 \circ
\mathbb{V}')^* ] \circ \Psi
\end{eqnarray}
where, $ \Psi = - \emph{\textbf{I}} \circ \mathbb{M} /\hbar $ is the
wave function.

In the above equation, the conservation of wave function is
influenced by the field potential, field strength, field source,
velocity, charge and mass etc. When the linear momentum $P_j$ is
equal approximately to 0, and the wave function is $ \Psi = \psi(r)
exp(-\emph{\textbf{I}}~ E_d t/\hbar )$, the above equation can be
simplified as
\begin{eqnarray}
0 = &&  [ (p_0 - \emph{\textbf{I}}~ E_d/c)^2 - (p_1)^2 - (p_2)^2 -
(p_3)^2 + q \hbar \lozenge^* \circ \mathbb{A}'^* ] \circ \psi (r)
\nonumber\\
 \approx && [ (qa'_0 - \emph{\textbf{I}}~ E_d/c) - (1/2dc) \left\{ (p_1)^2 + (p_2)^2
+ (p_3)^2 \right\} + (q\hbar/2dc) (\lozenge^* \circ \mathbb{A}'^*) ]
\circ \psi (r)
\end{eqnarray}
where, $E_d$ is the 'energy'; $(q \hbar/2d)(\lozenge^* \circ
\mathbb{A}'^*)$ is the interplay term of the
electromagnetic-gravitational subfield with the spin $(q \hbar/2d)$.

Limited within certain conditions, Eq.(11) of the
electromagnetic-gravitational field in the octonion space can deduce
the Dirac and Schrodinger equations and their conclusions about
interplays of dark matter and ordinary matter, including the spin
and magnetic moment etc.

\subsection{Intermediate particle equations}

In the octonion space, the electromagnetic-gravitational subfield
and electromagnetic-electromagnetic subfield are generated by the
physical object $M$ which owns rotation and charge. The E current
and G current of the intermediate particle $N(d,q)$ are $(S_0^e ,
S_1^e , S_2^e , S_3^e)$ and $(S_0^g , S_1^g , S_2^g , S_3^g)$
respectively. The $N(d,q)$ is the mixture of the intermediate
particles $\gamma_e^g$ (the photon) and $\gamma_e^e$ . When
$\mathbb{T} = 0$, the wave equation of particle $N(d,q)$ which moves
around $M$ is
\begin{eqnarray}
(\mathbb{W}/\alpha + \hbar \lozenge)^* \circ (\mathbb{B}/\hbar) = 0
\end{eqnarray}

When the $\mathbb{B}$ is small, and the sum of last two terms is
equal approximately to zero, the above equation can be written as
follows
\begin{eqnarray}
\mathbb{W}/\alpha = && d \mathbb{V}' + q \mathbb{A}'
\nonumber\\
= && (qa'_0 + dv'_0) + \emph{\textbf{i}}_1 (qa'_1 + dv'_1) +
\emph{\textbf{i}}_2 (qa'_2 + dv'_2) + \emph{\textbf{i}}_3 (qa'_3 +
dv'_3)
\nonumber\\
&& + \emph{\textbf{I}}_0 (qA'_0 + dV'_0) + \emph{\textbf{I}}_1
(qA'_1 + dV'_1) + \emph{\textbf{I}}_2 (qA'_2 + dV'_2) +
\emph{\textbf{I}}_3 (qA'_3 + dV'_3)
\nonumber\\
= && p_0 + \emph{\textbf{i}}_1 p_1 + \emph{\textbf{i}}_2 p_2 +
\emph{\textbf{i}}_3 p_3 + \emph{\textbf{I}}_0 P_0 +
\emph{\textbf{I}}_1 P_1 + \emph{\textbf{I}}_2 P_2 +
\emph{\textbf{I}}_3 P_3 \nonumber
\end{eqnarray}
where, $\mathbb{A}' = (c k_{rx} k_b \mu_e^g/\mu_g^g) \lozenge^*
\circ (\mathbb{X} \circ \emph{\textbf{I}}_0 ) = a'_0 +
\emph{\textbf{i}}_1 a'_1 + \emph{\textbf{i}}_2 a'_2 +
\emph{\textbf{i}}_3 a'_3 + \emph{\textbf{I}}_0 A'_0 +
\emph{\textbf{I}}_1 A'_1 + \emph{\textbf{I}}_2 A'_2 +
\emph{\textbf{I}}_3 A'_3 $; $S_0^g = qc , S_0^e = dc$; $\mathbb{V}'
= (k_b \mu_e^e/\mu_g^g) \mathbb{V} = v'_0 + \emph{\textbf{i}}_1 v'_1
+ \emph{\textbf{i}}_2 v'_2 + \emph{\textbf{i}}_3 v'_3 +
\emph{\textbf{I}}_0 V'_0 + \emph{\textbf{I}}_1 V'_1 +
\emph{\textbf{I}}_2 V'_2 + \emph{\textbf{I}}_3 V'_3 $ ; $p_j = qa'_j
+ dv'_j ~; P_j = qA'_j + dV'_j ~; j = 0, 1, 2, 3$.

Therefore
\begin{eqnarray}
0 = && ( \mathbb{W}/\alpha + \hbar \lozenge)^* \circ \left\{ (
\mathbb{W}/\alpha + \hbar \lozenge)^*  \circ \Psi \right\}
\nonumber\\
\approx &&  [ (p_0 + \hbar \partial_{g0})^2 - (p_1 + \hbar
\partial_{g1})^2 - (p_2 + \hbar \partial_{g2})^2 - (p_3 + \hbar \partial_{g3})^2
- (P_0 + \hbar \partial_{e0})^2
\nonumber\\
&& - (P_1 + \hbar \partial_{e1})^2 - (P_2 + \hbar \partial_{e2})^2 -
(P_3 + \hbar \partial_{e3})^2 + q \hbar \lozenge^* \circ
\mathbb{A}'^* + d \hbar \lozenge^* \circ ( \emph{\textbf{I}}_0 \circ
\mathbb{V}')^* ] \circ \Psi
\end{eqnarray}
where, $ \Psi = \mathbb{B} /\hbar $ is the wave function; $(q\hbar
/d)(\lozenge^* \circ \mathbb{A}'^*)$ is the interplay term of the
electromagnetic-electromagnetic subfield with the spin $(q\hbar/d)$.

The above equation can be used to describe the quantum
characteristics of intermediate particles which possess the spin
$(q\hbar/d)$, E charge and G charge. Limited within certain
conditions, Eq.(13) of the electromagnetic-gravitational field in
the octonion space can deduce the wave equation and its conclusions
about the interplays of ordinary matter and dark matter.

\subsection{Dirac-like equation}

In the octonion space, the electromagnetic-gravitational subfield
and electromagnetic-electromagnetic subfield are generated by the
physical object $M$ which owns rotation and charge. The E current
and G current of the intermediate particle $N(d,q)$ are $(S_0^e ,
S_1^e , S_2^e , S_3^e)$ and $(S_0^g , S_1^g , S_2^g , S_3^g)$
respectively. The $N(d,q)$ is the mixture of the intermediate
particles $\gamma_e^e$ and $\gamma_e^g$ . When $\mathbb{T} = 0$, the
wave equation of particle $N(d,q)$ which moves around $M$ is
\begin{eqnarray}
(\mathbb{W}/\alpha + \hbar \lozenge)^* \circ (\mathbb{B}/\hbar) = 0
\end{eqnarray}

When the energy $\mathbb{W} = 0$, the Dirac-like equation can be
attained from the above equation
\begin{eqnarray}
\hbar \lozenge^* \circ (\mathbb{B}/\hbar) = 0
\end{eqnarray}

From the above equation, we can conclude that intermediate particles
possess spin $(q\hbar/d)$ with no G charge nor E charge, and obtain
the corresponding quantum equation. Limited within certain
conditions, Eq.(13) of the electromagnetic-gravitational field in
the octonion space can deduce Dirac-like equation and its
conclusions about the interplays of dark matter and ordinary matter.

\section{Quantization of dark matter B}

In the electromagnetic-gravitational field, we can research the
interplays among the gravitational-gravitational subfield (ordinary
matter field) with gravitational-electromagnetic subfield (dark
matter field), and describe their quantum properties of the field
source particle and intermediate particle.

\subsection{Dirac and Schrodinger equations}

In the octonion space, the gravitational-gravitational subfield and
gravitational-electromagnetic subfield are generated by the physical
object $M$ which owns rotation and charge. The E momentum and G
momentum of the field source particle $N(m, e)$ are $(s_0^e , s_1^e
, s_2^e , s_3^e)$ and $(s_0^g , s_1^g , s_2^g , s_3^g)$
respectively. The $N(m, e)$ is the mixture of the general charges
$Q_g^g$ (the mass, $m$) and $Q_g^e$ (the dark matter charge, $e$).
When $\mathbb{U} = 0$, the wave equation of the particle $N(m,e)$
which moves around $M$ is
\begin{eqnarray}
(\mathbb{W}/\alpha + \hbar \lozenge)^* \circ (- \emph{\textbf{I}}
\circ \mathbb{M}/\hbar) = 0
\end{eqnarray}

Because of $ | s_0^g | \gg |  s_i^g | $ and $ | s_0^e | \gg | s_i^e
| $ , then $(i = 1, 2, 3)$
\begin{eqnarray}
\mathbb{W} && = (\mathbb{B} + \alpha \lozenge)^* \circ \left\{ (
\mathbb{R} + k_{rx} \mathbb{X} ) \circ \mathbb{P} \right\}
\nonumber\\
&& \approx \alpha \lozenge^* \circ \left\{ ( \mathbb{R} + k_{rx}
\mathbb{X} ) \circ (\mu_g^g s_0^g + \mu_g^e \emph{\textbf{I}}_0
s_0^e) \right\}/\mu_g^g + \mathbb{B}^* \circ \mathbb{M}
\nonumber\\
&& \approx \alpha (m \mathbb{V}' + e \mathbb{A}') + \mathbb{B}^*
\circ \mathbb{M} + \alpha [ \mu_g^e s_0^e \left\{ \lozenge^* \circ (
\mathbb{R} \circ \emph{\textbf{I}}_0 )\right\} + \mu_g^g k_{rx}
s_0^g \mathbb{A}] / \mu_g^g \nonumber
\end{eqnarray}
where, $\mathbb{A}' = (c k_{rx} \mu_g^e/\mu_g^g) \lozenge^* \circ
(\mathbb{X} \circ \emph{\textbf{I}}_0 ) = a'_0 + \emph{\textbf{i}}_1
a'_1 + \emph{\textbf{i}}_2 a'_2 + \emph{\textbf{i}}_3 a'_3 +
\emph{\textbf{I}}_0 A'_0 + \emph{\textbf{I}}_1 A'_1 +
\emph{\textbf{I}}_2 A'_2 + \emph{\textbf{I}}_3 A'_3 $; $s_0^g = mc ,
s_0^e = ec$; $\mathbb{V} = v_0 + \emph{\textbf{i}}_1 v_1 +
\emph{\textbf{i}}_2 v_2 + \emph{\textbf{i}}_3 v_3 +
\emph{\textbf{I}}_0 V_0 + \emph{\textbf{I}}_1 V_1 +
\emph{\textbf{I}}_2 V_2 + \emph{\textbf{I}}_3 V_3 $ ; $e $ is the
$Q_g^e$, $m$ is the $Q_g^g$ .

When the $\mathbb{B}$ is small, and the sum of last two terms is
equal approximately to zero, the above equation can be written as
follows
\begin{eqnarray}
\mathbb{W}/\alpha = && m \mathbb{V} + e \mathbb{A}'
\nonumber\\
= && (ea'_0 + mv_0) + \emph{\textbf{i}}_1 (ea'_1 + mv_1) +
\emph{\textbf{i}}_2 (ea'_2 + mv_2) + \emph{\textbf{i}}_3 (ea'_3 +
mv_3)
\nonumber\\
&& + \emph{\textbf{I}}_0 (eA'_0 + mV_0) + \emph{\textbf{I}}_1 (eA'_1
+ mV_1) + \emph{\textbf{I}}_2 (eA'_2 + mV_2) + \emph{\textbf{I}}_3
(eA'_3 + mV_3)
\nonumber\\
= && p_0 + \emph{\textbf{i}}_1 p_1 + \emph{\textbf{i}}_2 p_2 +
\emph{\textbf{i}}_3 p_3 + \emph{\textbf{I}}_0 P_0 +
\emph{\textbf{I}}_1 P_1 + \emph{\textbf{I}}_2 P_2 +
\emph{\textbf{I}}_3 P_3 \nonumber
\end{eqnarray}
where, $p_j = ea'_j + mv_j ~; P_j = eA'_j + mV_j ~; j = 0, 1, 2, 3$.

Therefore
\begin{eqnarray}
0 = && ( \mathbb{W}/\alpha + \hbar \lozenge)^* \circ \left\{ (
\mathbb{W}/\alpha + \hbar \lozenge)^*  \circ \Psi \right\}
\nonumber\\
\approx &&  [ (p_0 + \hbar \partial_{g0})^2 - (p_1 + \hbar
\partial_{g1})^2 - (p_2 + \hbar \partial_{g2})^2 - (p_3 + \hbar \partial_{g3})^2
- (P_0 + \hbar \partial_{e0})^2
\nonumber\\
&& - (P_1 + \hbar \partial_{e1})^2 - (P_2 + \hbar \partial_{e2})^2 -
(P_3 + \hbar \partial_{e3})^2 + e \hbar \lozenge^* \circ
\mathbb{A}'^* + m \hbar \lozenge^* \circ ( \emph{\textbf{I}}_0 \circ
\mathbb{V})^* ] \circ \Psi
\end{eqnarray}
where, $ \Psi = - \emph{\textbf{I}} \circ \mathbb{M} /\hbar $ is the
wave function.

In the above equation, the conservation of wave function is
influenced by the field potential, field strength, field source,
velocity, charge and mass etc. When the linear momentum $P_j$ is
equal approximately to 0, and the wave function is $ \Psi = \psi(r)
exp(-\emph{\textbf{I}}~ E t/\hbar )$, the above equation can be
simplified as
\begin{eqnarray}
0 = &&  [ (p_0 - \emph{\textbf{I}}~ E/c)^2 - (p_1)^2 - (p_2)^2 -
(p_3)^2 + e \hbar \lozenge^* \circ \mathbb{A}'^* ] \circ \psi (r)
\nonumber\\
\approx && [ (ea'_0 - \emph{\textbf{I}}~ E/c) - (1/2mc) \left\{
(p_1)^2 + (p_2)^2 + (p_3)^2 \right\} + (e\hbar/2mc) (\lozenge^*
\circ \mathbb{A}'^*) ] \circ \psi (r)
\end{eqnarray}
where, $E$ is the energy; $(e \hbar/2m)(\lozenge^* \circ
\mathbb{A}'^*)$ is the interplay term of the
gravitational-electromagnetic subfield with the spin $(e \hbar/2m)$.

Limited within certain conditions, Eq.(11) of the
electromagnetic-gravitational field in the octonion space can deduce
the Dirac and Schrodinger equations and their conclusions about
interplays of dark matter and ordinary matter, including the 'spin'
and 'magnetic moment' etc.

\subsection{Intermediate particle equations}

In the octonion space, the gravitational-gravitational subfield and
gravitational-electromagnetic subfield are generated by the physical
object $M$ which owns rotation and charge. The E momentum and G
momentum of the field source particle $N(m, e)$ are $(s_0^e , s_1^e
, s_2^e , s_3^e)$ and $(s_0^g , s_1^g , s_2^g , s_3^g)$
respectively. The $N(m, e)$ is the mixture of the intermediate
particles $\gamma_g^g$ and $\gamma_g^e$ . When $\mathbb{T} = 0$, the
wave equation of the particle $N(m,e)$ which moves around $M$ is
\begin{eqnarray}
(\mathbb{W}/\alpha + \hbar \lozenge)^* \circ (\mathbb{B}/\hbar) = 0
\end{eqnarray}

When the $\mathbb{B}$ is small, and the sum of last two terms is
equal approximately to zero, the above equation can be written as
follows
\begin{eqnarray}
\mathbb{W}/\alpha = && m \mathbb{V} + e \mathbb{A}'
\nonumber\\
= && (ea'_0 + mv_0) + \emph{\textbf{i}}_1 (ea'_1 + mv_1) +
\emph{\textbf{i}}_2 (ea'_2 + mv_2) + \emph{\textbf{i}}_3 (ea'_3 +
mv_3)
\nonumber\\
&& + \emph{\textbf{I}}_0 (eA'_0 + mV_0) + \emph{\textbf{I}}_1 (eA'_1
+ mV_1) + \emph{\textbf{I}}_2 (eA'_2 + mV_2) + \emph{\textbf{I}}_3
(eA'_3 + mV_3)
\nonumber\\
= && p_0 + \emph{\textbf{i}}_1 p_1 + \emph{\textbf{i}}_2 p_2 +
\emph{\textbf{i}}_3 p_3 + \emph{\textbf{I}}_0 P_0 +
\emph{\textbf{I}}_1 P_1 + \emph{\textbf{I}}_2 P_2 +
\emph{\textbf{I}}_3 P_3 \nonumber
\end{eqnarray}
where, $\mathbb{A}' = (c k_{rx} \mu_g^e/\mu_g^g) \lozenge^* \circ
(\mathbb{X} \circ \emph{\textbf{I}}_0 ) = a'_0 + \emph{\textbf{i}}_1
a'_1 + \emph{\textbf{i}}_2 a'_2 + \emph{\textbf{i}}_3 a'_3 +
\emph{\textbf{I}}_0 A'_0 + \emph{\textbf{I}}_1 A'_1 +
\emph{\textbf{I}}_2 A'_2 + \emph{\textbf{I}}_3 A'_3 $; $s_0^g = mc ,
s_0^e = ec$; $\mathbb{V} = v_0 + \emph{\textbf{i}}_1 v_1 +
\emph{\textbf{i}}_2 v_2 + \emph{\textbf{i}}_3 v_3 +
\emph{\textbf{I}}_0 V_0 + \emph{\textbf{I}}_1 V_1 +
\emph{\textbf{I}}_2 V_2 + \emph{\textbf{I}}_3 V_3 $ ; $p_j = ea'_j +
mv_j ~; P_j = eA'_j + mV_j ~; j = 0, 1, 2, 3$; $e $ is the $Q_g^e$,
$m$ is the $Q_g^g$ .

Therefore
\begin{eqnarray}
0 = && ( \mathbb{W}/\alpha + \hbar \lozenge)^* \circ \left\{ (
\mathbb{W}/\alpha + \hbar \lozenge)^*  \circ \Psi \right\}
\nonumber\\
\approx &&  [ (p_0 + \hbar \partial_{g0})^2 - (p_1 + \hbar
\partial_{g1})^2 - (p_2 + \hbar \partial_{g2})^2 - (p_3 + \hbar \partial_{g3})^2
- (P_0 + \hbar \partial_{e0})^2
\nonumber\\
&& - (P_1 + \hbar \partial_{e1})^2 - (P_2 + \hbar \partial_{e2})^2 -
(P_3 + \hbar \partial_{e3})^2 + e \hbar \lozenge^* \circ
\mathbb{A}'^* + m \hbar \lozenge^* \circ ( \emph{\textbf{I}}_0 \circ
\mathbb{V})^* ] \circ \Psi
\end{eqnarray}
where, $ \Psi = \mathbb{B} /\hbar $ is the wave function; $(e\hbar
/m)(\lozenge^* \circ \mathbb{A}'^*)$ is the interplay term of the
gravitational-electromagnetic subfield with the 'spin' $(e\hbar/m)$.

The above equation can be used to describe the quantum
characteristics of intermediate particles which possess the spin
$(e\hbar/m)$, E mass and G mass. Limited within certain conditions,
Eq.(13) of the electromagnetic-gravitational field in the octonion
space can deduce the wave equation and its conclusions about the
interplays of ordinary matter and dark matter.

\subsection{Dirac-like equation}

In the octonion space, the gravitational-gravitational subfield and
gravitational-electromagnetic subfield are generated by the physical
object $M$ which owns rotation and charge. The E momentum and G
momentum of the field source particle $N(m, e)$ are $(s_0^e , s_1^e
, s_2^e , s_3^e)$ and $(s_0^g , s_1^g , s_2^g , s_3^g)$
respectively. The $N(m, e)$ is the mixture of the intermediate
particles $\gamma_g^g$ and $\gamma_g^e$ . When $\mathbb{T} = 0$, the
wave equation of the particle $N(m,e)$ which moves around $M$ is
\begin{eqnarray}
(\mathbb{W}/\alpha + \hbar \lozenge)^* \circ (\mathbb{B}/\hbar) = 0
\end{eqnarray}

When the energy $\mathbb{W} = 0$, the Dirac-like equation can be
attained from the above equation
\begin{eqnarray}
\hbar \lozenge^* \circ (\mathbb{B}/\hbar) = 0
\end{eqnarray}

From the above equation, we can conclude that intermediate particles
possess the spin $(e\hbar/m)$ with no G mass nor E mass, and obtain
the corresponding quantum equation. Limited within certain
conditions, Eq.(13) of the electromagnetic-gravitational field in
the octonion space can deduce Dirac-like equation and its
conclusions about the interplays of the dark matter and the ordinary
matter.

\section{CONCLUSIONS}

The paper describes the quantization theory of electromagnetic and
gravitational interactions and dark matter, including the Dirac
equation, Schrodinger equation, and Dirac-like equation etc. And
there exist the quantum interplays of ordinary matter and dark
matter in the octonion space.

In the electromagnetic-gravitational subfield (electromagnetic
field) and gravitational-gravitational subfield (gravitational
field), we extend Dirac and Schrodinger equations of the field
source particles, and deduce the Dirac-like equation of intermediate
particles. It predicts that there exists one sort of field source
particle (electric charge and mass), which possesses the spin
$(q\hbar/2m)$. And it also predicts that there exists one kind of
intermediate particle, which is the mixture of the $\gamma_e^g$
(photon) and $\gamma_g^g$, and may possess the spin $(q\hbar/m)$
with no G mass nor G charge.

In the electromagnetic-gravitational subfield (electromagnetic
field) and electromagnetic-electromagnetic subfield (dark matter
field), we infer the Dirac equation and Schrodinger equation of the
field source particles, and deduce the Dirac-like equation of
intermediate particles. It predicts that the dark matter field has
one sort of field source particle (electric charge and E charge),
which possesses the spin $(q\hbar/2d)$. And it also predicts that
the dark matter field has one sort of the intermediate particle,
which is the mixture of the $\gamma_e^g$ (photon) and $\gamma_e^e$,
and may possess the spin $(q\hbar/d)$ with no E charge nor G charge.

In the gravitational-gravitational subfield (gravitational field)
and gravitational-electromagnetic subfield (dark matter field), we
extend Dirac and Schrodinger equations of the field source
particles, and deduce the Dirac-like equation of intermediate
particles. It predicts that there exists one sort of field source
particle (G mass and E mass), which possesses the spin
$(e\hbar/2m)$. And it also predicts that there exists one kind of
intermediate particle, which is the mixture of the $\gamma_g^e$ and
$\gamma_g^g$, and may possess the spin $(e\hbar/m)$ with no G mass
nor E mass.

\begin{acknowledgments}
This project was supported partly by the National Natural Science
Foundation of China under grant number 60677039, Science \&
Technology Department of Fujian Province of China under grant number
2005HZ1020 and 2006H0092, and Xiamen Science \& Technology Bureau of
China under grant number 3502Z20055011.
\end{acknowledgments}


\begin{references}

\bibitem{bosma} A. Bosma, \emph{Celestial Mechanics \& Dynamical Astronomy}, \textbf{72}, 69 (1998).

\bibitem{rubin} V. C. Rubin, A. H. Waterman and J. D. P. Kenney, \emph{Astronomical
Journal}, \textbf{118}, 236 (1999).

\bibitem{weng} Z. Weng, arXiv:physics/0612102.

\bibitem{adler} S. L. Adler, \emph{Quaternionic Quantum Mechanics and Quantum Fields},
(Oxford University Press, New York, 1995).

\bibitem{wu} N. Wu, \emph{Commun. Theor. Phys.}, \textbf{45}, 452 (2006).

\bibitem{boulanger} N. Boulanger, F. Buisseret, P. Spindel, \emph{Phys. Rev. D}, \textbf{74}, 125014 (2006).

\bibitem{singh} D. Singh, N. Mobed, G. Papini, \emph{Phys. Rev. Lett.}, \textbf{97}, 041101 (2006).

\bibitem{lomont} J. S. Lomont, \emph{Phys. Rev.}, \textbf{111}, 1710 (1958).

\bibitem{moses} H. E. Moses, \emph{Phys. Rev.}, \textbf{113}, 1670 (1959).

\bibitem{corvino} G. Corvino, G. Montani, \emph{Mod. Phys. Lett. A}, \textbf{19}, 2777 (2004).

\bibitem{oaknin} D. H. Oaknin, A. Zhitnitsky, \emph{Phys. Rev. D}, \textbf{71}, 023519 (2005).

\bibitem{galley} C. R. Galley, B. L. Hu, S.-Y. Lin, \emph{Phys. Rev. D}, \textbf{74}, 024017 (2006).

\end{references}
\end{document}